\documentclass[12pt]{iopart}
\usepackage{graphicx}
\usepackage[bookmarks,dvips,pdfhighlight=/O,pdfstartview=FitH]{hyperref}

\usepackage{iopams}  

\newcommand{\beq}{\begin{equation}}
\newcommand{\eeq}{\end{equation}}
\newcommand{\bea}{\begin{eqnarray}}
\newcommand{\eea}{\end{eqnarray}}
\newcommand{\bce}{\begin{center}}
\newcommand{\ece}{\end{center}}

\newcommand{\dd}{\mathrm{d}}
\def\lsim{\mathrel{\rlap{\lower4pt\hbox{\hskip1pt$\sim$}}
    \raise1pt\hbox{$<$}}}         
\def\gsim{\mathrel{\rlap{\lower4pt\hbox{\hskip1pt$\sim$}}
    \raise1pt\hbox{$>$}}}         

\begin{document}

\title{Thermal Dileptons at LHC}

\author{H van Hees and R Rapp} 

\address{Cyclotron Institute and Physics Department, 
Texas A\&M University, College Station, Texas 77843-3366,
  U.S.A.}  

\date{June 27, 2007}

\begin{abstract}
  We predict dilepton invariant-mass spectra for central 5.5~ATeV Pb-Pb
  collisions at LHC. Hadronic emission in the low-mass region is
  calculated using in-medium spectral functions of light vector mesons
  within hadronic many-body theory. In the intermediate-mass region
  thermal radiation from the Quark-Gluon Plasma, evaluated
  perturbatively with hard-thermal loop corrections, takes over. An
  important source over the entire mass range are decays of correlated
  open-charm hadrons, rendering the nuclear modification of charm and
  bottom spectra a critical ingredient.
\end{abstract}

Due to their penetrating nature, electromagnetic probes (dileptons and
photons) are an invaluable tool to investigate direct radiation from the
hot/dense matter created in heavy-ion collisions.  At low
invariant-mass, $M$$\leq$1~GeV, the main source of dileptons is the
decay of the light vector mesons, $\rho$, $\omega$ and $\phi$, giving
unique access to their in-medium spectral properties, most prominently
for the short-lived $\rho$ meson. If the chiral properties of the
$\rho$-meson can be understood theoretically, dilepton spectra can serve
as a signal for the restoration of chiral symmetry at high temperatures
and densities.

We employ medium-modified vector-meson spectral functions in hot/dense
matter following from hadronic many-body theory, phenomenologically
constrained by vacuum $\pi\pi$ scattering, decay branching ratios for
baryonic and mesonic resonances, photo-absorption cross sections on
nucleons and nuclei, etc.~\cite{Rapp:1999us}. The resulting spectral
functions, especially for the $\rho$ meson, exhibit large broadening
with little mass shift, with baryonic interactions as the prevalent
agent, especially in the mass region below the resonance peaks. Note
that $CP$ invariance of strong interactions implies equal interactions
with baryons and antibaryons. Thus, even in a net-baryon free
environment, the $\rho$ resonance essentially ``melts'' around the
expected phase transition temperature, $T_c$$\simeq$180~MeV.  Other
sources of thermal dileptons taken into account are (i) four-pion type
annihilation in the hadronic phase (augmented by chiral
vector-axialvector mixing)~\cite{vanHees:2006ng}, which takes over the
resonance contributions at intermediate mass, and (ii) radiation from
the Quark-Gluon Plasma (QGP), computed within hard-thermal loop improved
perturbation theory for in-medium $q$-$\bar{q}$ annihilation.

Thermal dilepton spectra are computed by evolving pertinent emission
rates over the time evolution of the medium in central 5.5~ATeV Pb-Pb
collisions.  To this end, we employ a cylindrical homogeneous thermal
fireball with isentropic expansion and a total entropy fixed by the
number of charged particles, which we estimate from a phenomenological
extrapolation to be $\dd N_{\mathrm{ch}}/\dd y$$\simeq$1400. We use an
ideal-gas equation of state (EoS) with massless gluons and $N_f$=2.5
quark flavors for the QGP, and a resonance gas for the hadronic EoS with
chemical freezeout at ($\mu_B^c$,$T_c$)=(2,180)~MeV (finite meson and
anti-/baryon chemical potentials are implemented to conserve the
particle ratios until thermal freezeout at
$T_{\mathrm{fo}}$$\simeq$100~MeV, with a mass-action law for short-lived
resonances).  We start the evolution in the QGP phase at initial time
$\tau_0$=0.17~fm/c, translating into $T_0$$\simeq$560~MeV.  The volume
expansion parameters are taken to resemble hydrodynamic simulations. A
standard mixed-phase construction connects QGP and hadronic phase at
$T_c$, and the total fireball lifetime is
$\tau_{\mathrm{fb}}$$\simeq$18~fm/c.
\begin{figure}
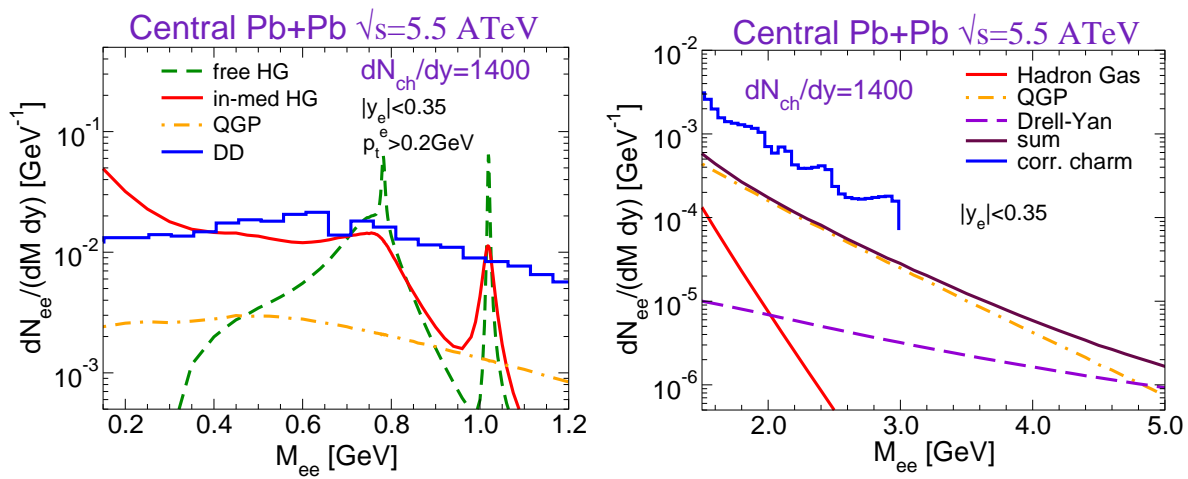

\begin{center}
\begin{minipage}{0.49\textwidth}
\includegraphics[width=\textwidth]{dNdMPb5500lm-Nch1400}
\end{minipage}\hspace*{1mm}
\begin{minipage}{0.49\textwidth}
\includegraphics[width=\textwidth]{dNdMPb5500im-Nch1400}
\end{minipage}
\end{center}
\caption{(Color online) Predictions for dilepton spectra in 
central 5.5~ATeV Pb-Pb collisions at LHC in the 
low- (left panel) and intermediate-mass region (right panel).}
\label{fig.1}
\end{figure}

As for non-thermal sources, we include primordial Drell-Yan annihilation
and decays of correlated charm pairs.  The latter are estimated by
scaling the spectrum at RHIC with a charm-cross section anticipated at
LHC, which implies somewhat softer charm spectra than expected for
primordial N-N collisions (and thus softer invariant-mass spectra).
We neglect contributions from jet-plasma interactions.

Our predictions are summarized in Fig.~\ref{fig.1}. At low mass 
thermal dileptons are dominated by hadronic radiation, with large
modifications due to in-medium vector-meson spectral functions. The QGP
contribution takes over at around $M$$\gtrsim$1.1~GeV. The yield from
correlated open-charm decays is comparable to hadronic emission already
at low mass, and dominant at intermediate mass. However, this result
will have to be scrutinized by including the nuclear modification
of heavy-quark spectra in the QGP (as well as analogous
contributions from correlated bottom decays). Also, larger values of $\dd
N_{\mathrm{ch}}/\dd y$ would help to outshine correlated open-charm
decays, at least at low mass.

{\it This work is supported by a U.S. NSF CAREER Award, grant no. PHY-0449489.}


\vspace*{0.2cm}
\begin{thebibliography}{99}

\bibitem{Rapp:1999us}
  R.~Rapp and J.~Wambach,
  Eur.\ Phys.\ J.\  A {\bf 6}, 415 (1999)

\bibitem{vanHees:2006ng}
  H.~van Hees and R.~Rapp,
  Phys.\ Rev.\ Lett.\  {\bf 97}, 102301 (2006)

\bibitem{vanHees:2006iv}
  H.~van Hees and R.~Rapp,
  Proceedings of the 22nd Winter Workshop on Nuclear Dynamics, EP Systema,
  Budapest, Hungary 2006, 91-98, 
  arXiv:hep-ph/0604269.

\end{thebibliography}
\end{document}